\documentclass[aps,prl,twocolumn,groupedaddress]{revtex4}

\usepackage{amsmath,amssymb}

\usepackage{graphicx}
\graphicspath{{Figures/}}
\DeclareGraphicsExtensions{.eps,.ps}

\begin{document}

\title{Quantum Necking in Stressed Metallic Nanowires}
\author{J.\ B\"urki,$^{1,3}$ Raymond E.\ Goldstein,$^{1,2}$ and C.\ A.\ 
Stafford$^{1}$}
\
\affiliation{$^{1}$Department of Physics and $^{2}$Program in Applied
Mathematics, University of Arizona, Tucson, AZ 85721 \\
$^{3}$Physikalisches Institut, Albert-Ludwigs-Universit\"at, D-79104
Freiburg, Germany} \date{\today}

\begin{abstract}
When a macroscopic metallic wire is subject to tensile stress, it
necks down smoothly as it elongates.  We show that nanowires with
radii comparable to the Fermi wavelength display remarkably different
behavior.  Using concepts from fluid dynamics, a PDE for nanowire
shape evolution is derived from a semiclassical energy functional that
includes electron-shell effects.  A rich dynamics involving movement
and interaction of kinks connecting locally stable radii is found, and
a new class of universal equilibrium shapes is predicted.
\end{abstract}


\maketitle

It has recently become possible to image metal nanowires with
subatomic resolution, and record their structural evolution in real
time, using transmission-electron microscopy
\cite{kondo97,kizuka98a,kondo00,rodrigues01,rodrigues02}, thus opening
the door to a world of nonlinear dynamical phenomena at the nanoscale.
Heretofore, most information about the structure and dynamics of
nanowires was obtained indirectly, from transport and cohesive
properties \cite{rubio96,stalder96,untiedt97,yanson99,yanson00}.
Classical molecular dynamics simulations
\cite{landman90,todorov96,sorensen98,tosatti98} provide important
insights into atomic structure and dynamics of nanowires, but cannot
explain the observed stability \cite{kondo97,kondo00,rodrigues02} of
long nanowires, which would be unstable classically under surface
tension \cite{chand81}.  Quantum molecular dynamics can in principle
study this stability, but has so far been limited to systems too small
to address such issues (e.g.  single or double chains
\cite{landman00,bahn01}).

\begin{figure}[b]  
\includegraphics[width=8.5cm, keepaspectratio]{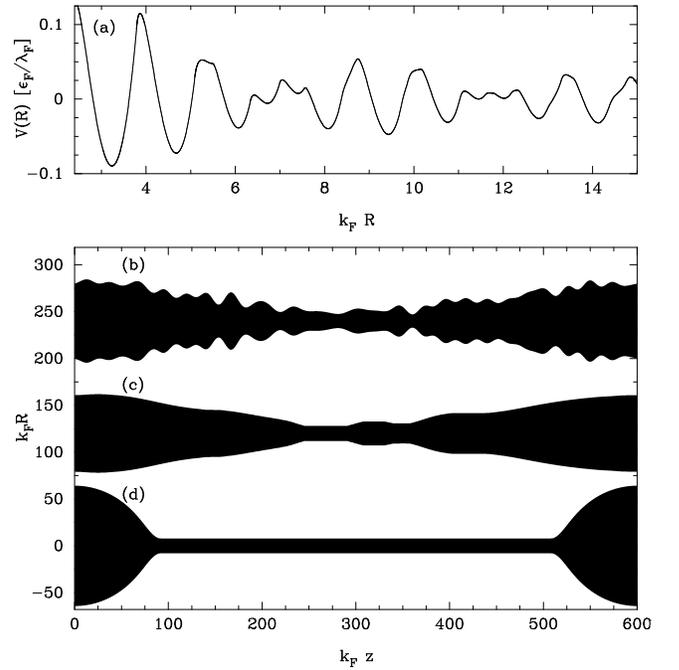}
\caption{\label{fig.shapes}
(a) Electron-shell potential $V(T,R)$ for
$T=0.002T_F$.  Here $\varepsilon_F$ and $\lambda_F$ are the Fermi
energy and wavelength, $T_F=\varepsilon_F/k_B$ and
$k_F=2\pi/\lambda_F$.  (b)--(d) Approach to equilibrium of an
initially random wire: (b) initial shape; (c) $\tau = 2\times 10^4$;
(d) $\tau = 3\times 10^7$, equilibrium wire with $G=12\,G_0$.  }
\end{figure}

\begin{figure}[b]  
\includegraphics[width=8.5cm, keepaspectratio]{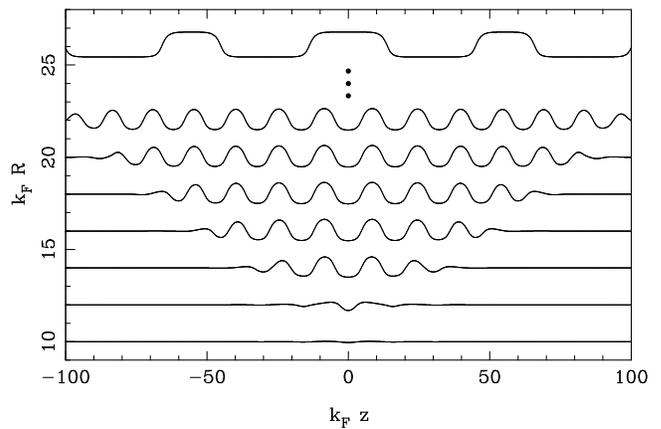}
\caption{\label{fig.prop} Propagation of an instability for a wire of
initial radius $k_F R = 10$, time progressing from bottom to top
($\Delta\tau=15$), with curves shifted for clarity.  The top curve
corresponds to a much later stage in the time evolution
($\tau=6\times10^4$).  }
\end{figure}

\begin{figure*}[t]  
\includegraphics[width=6.1in]{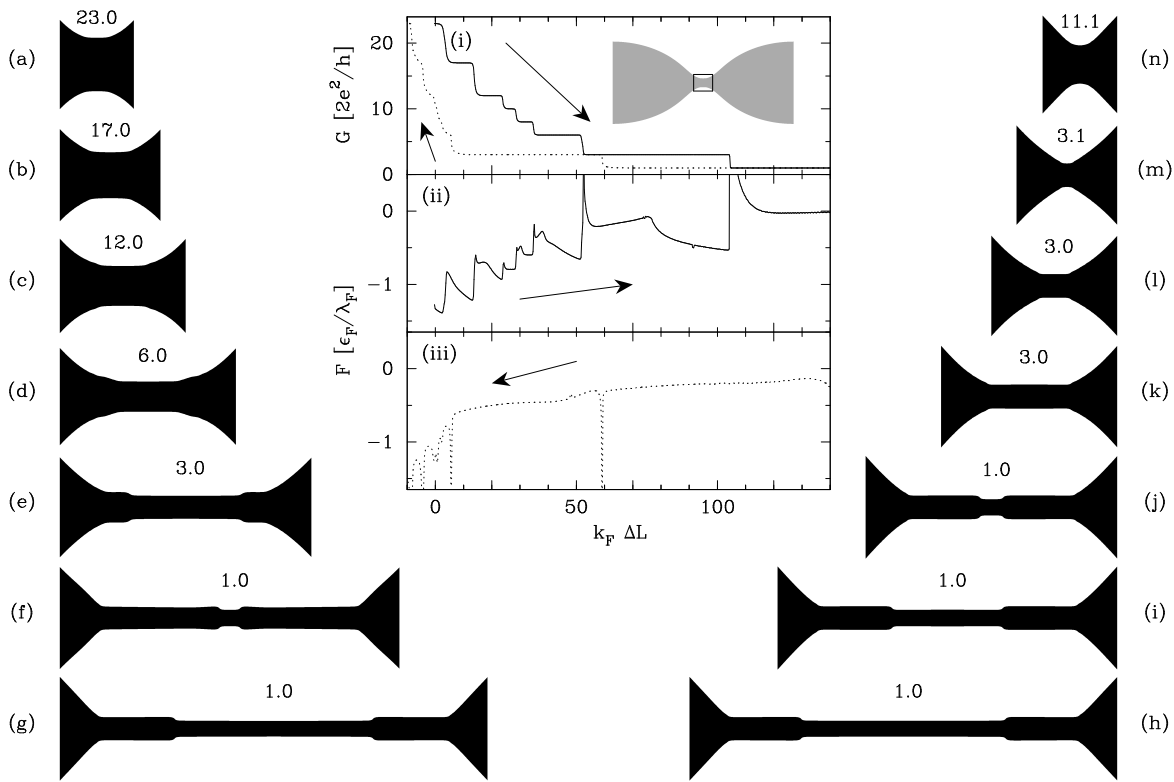}
\caption{\label{fig.cycle}\label{fig.F+G} 
Shape of a nanowire connecting two quasi-macroscopic leads during 
elongation (left column, a-g) and subsequent compression 
(right column, h-n) \cite{movies}.
For comparison, the entire system of length $L=280\,k_F^{-1}+\Delta L$ 
is shown in inset (i), the small square indicating the nanowire 
shown in image (a).  
Adjacent wires have the same length, from top to bottom: 
$k_F\Delta L=0, 10, 20, 40, 70, 105, 140$.  
Above each wire is its conductance in units of $G_0$.  
Central Inset: Conductance and force as a function of $\Delta L$: 
(i) Conductance during elongation (solid line) and compression 
(dotted line); (ii) force during elongation; 
(iii) force during compression.  }
\vspace*{-3mm}
\end{figure*}

By contrast, a nanoscale free-electron model
\cite{stafford97a,stafford99,kassubek01,zhang02} has proven successful
in explaining the quantum suppression of the Rayleigh instability
\cite{kassubek01}.  Here we extend this approach by developing a {\it
dynamics} for a free-electron nanowire under the assumption that
surface diffusion is a dominant process.  This dynamics, of instrinsic
interest in nonlinear science, explicitly includes quantum-size
effects and allows the study of a broad range of dynamical processes
(see Figs.\ \ref{fig.shapes}--\ref{fig.F+G}): Approach to equilibrium
of a nanowire, propagation of an instability, and evolution of a wire
under elongation/compression.  Figure\ \ref{fig.shapes} shows how an
initially random nanowire necks down to a universal shape, consisting
of a central cylinder connected to contacts having the form of
unduloids of revolution, providing a dynamical mechanism for the
nanofabrication technique invented by Kondo and Takayanagi
\cite{kondo97}.  Figure\ \ref{fig.F+G} compares well to early
experiments on gold nanocontacts \cite{rubio96,stalder96}, showing
perfect correlation between force and conductance with steep
conductance steps, and a hysteresis between elongation and
compression.

The model of a nanowire consists of free electrons confined to an
axisymmetric wire by a hard-wall potential \cite{stafford97a}.  The
wire radius is $R(z,t)$ in the interval $[0,L]$ along the $z$-axis,
and periodic boundary conditions \cite{com.leads} are used to extend
$R(z)$.  In contrast to the classical case \cite{chand81} for which
all nonaxisymmetric distortions raise the energy, Jahn-Teller
distortions that break axisymmetry of nanowires {\it do} lower the
energy in a few cases.  However,
the deepest energetic minima occur for axisymmetric wires \cite{nonaxi}, 
and axisymmetry,
once present, is preserved by the diffusion equation, so that
Jahn-Teller distortions are most likely suppressed dynamically.

In the spirit of the Born-Oppenheimer approximation, the total energy
of the nanowire is taken to be the electronic energy.  Since we are
dealing with an open system of electrons, the grand-canonical
potential is used, and can be separated into Weyl and mesoscopic
contributions \cite{strutinsky,stafford99,brack97}.  Dropping the
volume contribution, which is assumed to be constant, the energy
functional is
\begin{equation}\label{eq.omega}  
\Omega\bigl[T,R(z)\bigr]=\sigma(T){\cal S}\bigl[R(z)\bigr]+\int_0^L
V\bigl(T,R(z)\bigr)\text{d}z,
\end{equation}  
where $\sigma(T)$ is the surface tension \cite{com.sigma}, ${\cal S}$
is the surface area of the wire, and $V$ is a mesoscopic
electron-shell potential whose integral gives the so-called ``shell
correction" to the energy (as used, for instance, in the study of
clusters \cite{yannouleas95}).  Higher-order terms
\cite{stafford99,brack97} proportional to the mean curvature, etc.,
are neglected.  $V$ can be expressed in terms of a Gutzwiller-type
trace formula
\begin{equation}\label{eq.gutzwiller}  
V(T,R) = \frac{2\varepsilon_F}{\pi}
\sum_{w=1}^{\infty}\sum_{v=2w}^{\infty} a_{vw}(T)
\frac{f_{vw}\cos\theta_{vw}}{v^2L_{vw}},
\end{equation}  
where the sum includes all classical periodic orbits $(v,w)$ in a disk
billiard \cite{brack97}, characterized by their number of vertices $v$
and winding number $w$, $L_{vw}=2vR\sin(\pi w/v)$ is the length of
orbit $(v,w)$, and $\theta_{vw}=k_FL_{vw}-3v\pi/2$.  The factor
$f_{vw}=1\text{ for } v=2w, 2$ otherwise, accounts for the invariance
under time-reversal symmetry of some orbits, and $a_{vw}(T) =
\tau_{vw}/\sinh{\tau_{vw}}$ ($\tau_{vw}=\pi k_FL_{vw}T/2T_F$) is a
temperature-dependent damping factor.  A similar trace formula was
previously derived \cite{kassubek01} for small axisymmetric
perturbations of a cylinder using semiclassical perturbation theory
\cite{ullmo96,creagh96,brack97}.  Here we point out that semiclassical
perturbation theory remains valid for large deformations, as long as
new classes of non-planar orbits can be neglected (adiabatic
approximation).

Eqs.\ (\ref{eq.omega}) and (\ref{eq.gutzwiller}) define an energy
functional which is general and simple enough to solve for nontrivial
nanowire geometries with a wide range of radii.  The multiple deep
minima of $V(R)$, shown in Fig.\ \ref{fig.shapes}(a), favor certain
magic radii \cite{yanson99} and suggest a multistable field theory
analogous to the sine-Gordon model.  We therefore anticipate solutions
consisting of cylindrical segments connected by kinks.

Rather than directly minimizing the energy 
functional~(\ref{eq.omega}), we derive an equation describing the 
nanowire surface dynamics which yields the approach to equilibrium 
as well as the onset of instability.  
The model has two main assumptions.  First,
in the thin wires considered, the majority of atoms are at the surface
and thus surface self-diffusion is the mechanism of ionic motion.
Second, under the Born-Oppenheimer approximation, the electronic
energy (\ref{eq.omega}) acts as a potential for the ions, and the
dynamics derive from ionic mass conservation:
\begin{equation}\label{eq.diffusion}  
\frac{\partial R(z,t)}{\partial t} =
-\frac{v_a}{R(z,t)}\frac{\partial}{\partial
z}\bigl[R(z,t)J_z(z,t)\bigr],
\end{equation}  
where $v_a=3\pi^2/k_F^3$ is the volume of an atom, and the
$z$-component of the surface current is given by Fick's law:
\begin{equation}\label{eq.current} 
J_z = -\frac{\rho_SD_S}{k_B T}\frac{1}{\sqrt{1+(R')^2}}
\frac{\partial\mu}{\partial z},
\end{equation}  
where $R'=\partial R/\partial z$, and $\rho_S$ and $D_S$ are the
surface density of ions and the surface self-diffusion coefficient,
respectively.  The precise value of $D_S$ for alkali metals is not
known, but it can be removed from the evolution equation by rescaling
time to the dimensionless variable $\tau=(\rho_SD_ST_F/T)t$.  The
chemical potential $\mu$ is obtained by calculating the change in the
energy (\ref{eq.omega}) with the addition of an atom at point $z_0$,
$\mu(z_0) \equiv \Omega\bigl[T,R(z)+A\delta(z-z_0)\bigr]
-\Omega\bigl[T,R(z)\bigr]$, where $A=v_a/2\pi R$ is chosen so that the
volume of an atom is added:
\begin{multline}\label{eq.mu}  
\mu(z) =
-\frac{2\varepsilon_F}{5}+\frac{3\pi\sigma}{k_F^3R(z)\sqrt{1+{R'}^2}}
\Bigl(1-\frac{RR''}{1+{R'}^2}\Bigr) \\
-\frac{3\varepsilon_F}{k_F^2R^2}\sum_{w,v}
\frac{a_{vw}f_{vw}}{v^2}\Biggl[\sin\theta_{vw}+
b_{vw}\frac{\cos\theta_{vw}}{k_FL_{vw}}\Biggr],
\end{multline}  
where $b_{vw}(T) = a_{vw}\cosh\tau_{vw}$.  Eq.\ (\ref{eq.diffusion})
is then solved numerically using an implicit scheme.  Our non-linear
dynamical model, Eqs.\ (\ref{eq.diffusion}--\ref{eq.mu}), differs from
previous studies of axisymmetric surface self-diffusion
\cite{coleman95, eggers98, bernoff98} by the inclusion of
electron-shell effects [2nd line of Eq.\ (\ref{eq.mu})], which
fundamentally alter the dynamics.

To explore the equilibrium shapes of nanowires, we considered several
random initial configurations of various lengths and widths, one of
which is shown in Fig.\ \ref{fig.shapes}(b).  Under surface
self-diffusion, the short-wavelength fluctuations are rapidly smoothed
out [Fig.\ \ref{fig.shapes}(c)], and several cylindrical segments
connected by ``kinks'' are visible.  On a much longer time-scale, the
wires neck down to a universal equilibrium shape consisting of a
nearly perfect cylindrical region with a radius near one of the minima
of $V(R)$, connected to thicker leads [Fig.\ \ref{fig.shapes}(d)].  In
all fifteen cases studied, the leads closely approximate Delaunay
unduloids of revolution.  The unduloid is a stationary configuration,
unstable under surface tension alone \cite{bernoff98}, but stabilized
in the present model by electron-shell effects in the connecting
cylinder.  The unduloids arise as superpositions of large numbers of
closely spaced kinks.  Equilibrium wires with conductance $G/G_0 = 1,
3, 6, 12, 17, 23, \dots$ were obtained, where $G_0=2e^2/h$ is the
conductance quantum.  Intermediate conductance values may occur more
rarely.

Other cases of interest are the propagation of an instability
\cite{rayleighprop} and the homogeneous-inhomogeneous transition in a
nanowire \cite{zhang02}.  In the former case, starting with a cylinder
of an unstable radius, a maximum in $V(R)$ [Fig.\ \ref{fig.shapes}(a)],
and adding a localized Gaussian deformation of small amplitude ($0.001
R$), we find that the perturbation decreases during an incubation
period and then the instability sets in with essentially a single
Fourier component corresponding to the most unstable mode
\cite{zhang02}, saturates at a finite amplitude, and propagates with
constant velocity (see Fig.\ \ref{fig.prop}).  A similar instability is
observed in the pearling of membranes \cite{pearling1}.  The classical
Rayleigh instability itself \cite{rayleighprop} propagates in much the
same way, but does not saturate, instead leading to breakup of the
cylinder.  Both the wavelength \cite{zhang02} and front velocity of
the instability are strongly dependent on the initial radius.  Once
the instability is fully developed, kink-antikink pairs start
annihilating, eventually leading to a shape consisting of a series of
cylinders of different radii [corresponding to neighboring minima in
$V(R)$, Fig.\ \ref{fig.shapes}(a)], connected by kinks.  Starting from
a more realistic initial condition with random perturbations, the
instability occurs at several places more or less simultaneously.

The dynamical model also allows for the study of the evolution of a
nanowire under strain.  The length $L$ of the wire is changed at an 
average rate $dL/dt$ in a sequence of infinitesimal steps.
In each step, the wire deforms elastically, i.e.\ the change of length is
distributed along the wire as if each of its slices were Hookean, with
spring constant proportional to its cross-sectional area.  The radius of
each slice is simultaneously changed so as to conserve its volume. 
Between steps, the wire evolves according to Eq.\ (\ref{eq.diffusion}). 
We start with an equilibrated unduloid of outer radius $k_F R = 99$,
representing a nanocontact between two quasi-macroscopic leads [see
Fig.\ \ref{fig.cycle} (inset)].  We then strain the wire by elongating
it at a constant speed $k_FdL/d\tau=0.1$ until it necks down to a
long, atomically thin wire [Fig.\ \ref{fig.cycle}(b-g)], at which point
we let it partially equilibrate [Fig.\ \ref{fig.cycle}(h)].
Subsequently, we compress the wire at the same rate until it reaches
its initial length [Fig.\ \ref{fig.cycle}(i-n)].  During elongation,
the necking of the wire occurs by nucleation of kink-antikink pairs at
the center with subsequent motion of these kinks toward the leads,
leaving a long, cylindrical wire in the middle. 
During compression, the remaining kink-antikink pairs annihilate, 
but new kinks do not emerge from the leads.
Similar necking was observed for various elongation rates $k_F dL/d\tau \leq
1$.  For much faster rates, structural relaxation due to surface
diffusion is suppressed.

These different processes lead to a substantial hysteresis between
elongation and compression, as can be seen by computing the electrical
conductance of the contact and the force applied to it.  The force is
obtained from the change in the energy (\ref{eq.omega}) with
elongation, $F=-\partial\Omega/\partial L$.  The conductance is
computed from the Landauer formula, using the adiabatic and WKB
approximations \cite{stafford97a} to compute the transmission
probabilities.  Both quantities are shown in the central inset of
Fig.\ \ref{fig.F+G}.  Compared to our previous results obtained without
structural relaxation \cite{stafford97a}, the conductance steps and
relaxations of the force are much steeper, in better agreement with
experiments \cite{rubio96}, and a substantial hysteresis is observed
in the conductance.  It has been argued \cite{rubio96,untiedt97} that
the abruptness of the conductance steps, and their perfect correlation
with the jumps in the force, rule out an electronic mechanism.
However, we have shown that this behavior arises naturally in a model
which takes proper account of electronic quantum-size effects on the
structure, even when atomistic effects are neglected.

A final interesting feature is the spikes in the force that occur at
the opening and closing of a channel, corresponding to the rapid
energy relaxation during the creation or annihilation of a
kink/antikink pair.  These spikes are suppressed, and the hysteresis
decreases, when the speed of deformation is increased, suggesting that
the spikes might be observable experimentally by decreasing the
elongation rate.  A failure of this model is the force during
compression: contrary to experiment \cite{rubio96,untiedt97}, the
force is attractive, suggesting that a resistance to compression due
to the ions needs to be added to the free-electron model.  Indeed, an
interesting question raised by this work is how discrete positive ions
would arrange themselves to accomodate the predicted shapes favored by
the conduction electrons.


\begin{acknowledgments}
This research was supported in part by NSF grants DMR0072703 and DMR0312028 
(CAS \& JB), and by NSF grant DMR9812526 (REG).  
JB also acknowledges support from the Swiss National Science Foundation and
grant SFB276 of the Deutsche Forschungsgemeinschaft.
\end{acknowledgments}

\bibliography{refs}

\begin{thebibliography}{35}
\bibitem[{com({\natexlab{b}})}]{com.sigma}
\bibinfo{howpublished}{The value $\sigma=\varepsilon_F k_F^2/80\pi$ was used
  \cite{stafford99} to model alkali metals. For noble metals, $\sigma$ is
  roughly twice as large.}
\end{thebibliography}

\end{document}